\documentclass[twocolumn,twoside,slac_two]{revtex4}

\usepackage{graphicx}
\usepackage{fancyhdr}
\newcommand{\G}{\Gamma}

\newcommand{\sT}{\sigma_{\rm T}}
\newcommand{\p}{^\prime}

\newcommand{\g}{\gamma}

\newcommand{\dD}{\delta_{\rm D}}
\newcommand{\up}{u^\prime}

\newcommand{\psim}{\lower.5ex\hbox{$\; \buildrel \propto \over\sim \;$}}
\newcommand{\lbar}{\lower.0ex\hbox{$\; \buildrel
{\lower0.0ex \hbox{-}} \over\lambda  \;$}}

\newcommand{\erg}{\mathrm{erg}}

\newcommand{\s}{\mathrm{s}}

\newcommand{\Hz}{\mathrm{Hz}}

\begin{document}

\title{A Physical Model for the Revised Blazar Sequence}


\author{Justin D.\ Finke}
\affiliation{U.S.\ Naval Research Laboratory, Code 7653, 4555 Overlook Ave.\ SW,
        Washington, DC,
        20375-5352
}

\begin{abstract}

The blazar sequence is reflected in a correlation of the peak
luminosity versus peak frequency of the synchrotron component of
blazars. This correlation has been considered one of the fundamental
pieces of evidence for the existence of a continuous sequence that
includes low-power BL Lacertae objects through high-power flat
spectrum radio quasars.  A correlation between the Compton dominance,
the ratio of the Compton to synchrotron luminosities, and the peak
synchrotron frequency is another piece of evidence for the existence
of the blazar sequence explored by Fossati et al. (1998).  Since that
time, however, it has essentially been ignored.  We explore this
correlation with a sample based on the second LAT AGN catalog (2LAC),
and show that is is particularly important, since it is independent of
redshift.  We reproduce the trends in our sample with a simple model
that includes synchrotron and Compton cooling in the slow- and
fast-cooling regimes, and angle-dependent effects.

\end{abstract}

\maketitle

\thispagestyle{fancy}


\section{Introduction}
\label{intro}

Active Galactic Nuclei (AGN) with relativistic jets pointed along our
line of sight are known collectively as blazars.  This includes those
with strong broad emission lines, Flat Spectrum Radio Quasars (FSRQs),
and those with weak or absent lines, known as BL Lacertae objects
\citep[BL Lacs;][]{marcha96}.  In general, the spectral energy
distributions (SEDs) of blazars have two basic components: a low
frequency component, peaking in the optical through X-rays, from
synchrotron emission, and a high frequency component, peaking in the
$\g$ rays, probably originating from Compton scattering of some seed
photon source, either internal (synchrotron self-Compton or SSC) or
external to the jet (external Compton or EC).

Aside from their classifications as FSRQs or BL Lacs from optical
spectra, \cite{abdo10_sed} subdivided them based on their synchrotron
peak.  They are considered high synchrotron-peaked (HSP) blazars if
their synchrotron peak $\nu^{sy}_{pk} > 10^{15}\ \Hz$; intermediate
synchrotron-peaked (ISP) blazars if $10^{14}\ \Hz < \nu^{sy}_{pk} <
10^{15}\ \Hz$; and low synchrotron-peaked (LSP) blazars if
$\nu^{sy}_{pk}<10^{14}\ \Hz$.  Almost all FSRQs are LSP blazars.  

\cite{fossati98} combined several blazar surveys and noticed an
anti-correlation between the luminosity at the synchrotron peak,
$L_{pk}^{sy}$, and the frequency of this peak, $\nu^{sy}_{pk}$.  They
also noticed anti-correlations between the 5 GHz luminosity $L_{5\
GHz}$ and $\nu^{sy}_{pk}$, the $\g$-ray luminosity and
$\nu^{sy}_{pk}$, and the $\g$-ray dominance (the ratio of the EGRET
$\g$-ray luminosity to the synchrotron peak luminosity) and
$\nu^{sy}_{pk}$.  \cite{ghisellini98} provided a physical explanation
for these correlations \citep[see also][]{boett02_seq}.  If the seed
photon source for external Compton scattering is the broad-line region
(BLR), and the BLR strength is correlated with the power injected into
electrons in the jet, one would expect that more luminous jets have
stronger broad emission lines and greater Compton cooling, and thus a
lower $\nu^{sy}_{pk}$.  As the power injected in electrons is reduced,
the broad line luminosity decreases, there are fewer seed photons for
Compton scattering, and consequently the peak synchrotron frequency
moves to higher frequencies.  This is also reflected in the lower
luminosity of the Compton-scattered component relative to the
synchrotron component as $\nu^{sy}_{pk}$ moves to higher frequencies.

The $L_{pk}^{sy}$--$\nu_{pk}^{sy}$ anti-correlation has been
questioned.  Using blazars from two large surveys, \cite{padovani03}
did not find any anti-correlation between $\nu^{sy}_{pk}$ and radio,
BLR, or jet power.  This work, however, has been criticized for its
relatively poor SED characterization \citep{ghisellini08_seq}.  Also,
the lack of sources in the upper right region of the
$L_{pk}^{sy}$--$\nu_{pk}^{sy}$ plot could be the result of a selection
effect \citep{giommi02,padovani02,giommi05,giommi11_selection}.  Since
a large fraction of BL Lac objects have entirely featureless optical
spectra, their redshifts, $z$, and hence luminosities, are impossible
to determine.  These could be extremely bright, distant BL Lacs that
would fill in the upper right region.  \cite{nieppola06} found no
correlation between the frequency and luminosity of the synchrotron
peaks for objects in the Mets\"ahovi Radio Observatory BL Lacertae
sample.  \cite{chen11} did find an anti-correlation, using sources
found in the LAT bright AGN sample \citep{abdo09_lbas,abdo10_sed}.  In
recent works, an ``L''-shape in the $L_{pk}^{sy}$--$\nu_{pk}^{sy}$
plot seems to have emerged, as lower luminosity and low-peaked sources
have been detected with more sensitive instruments
\citep{meyer11,giommi12}.  \cite{nieppola06} found more of a ``V''
shape, although their plot did not include FSRQs; if high luminosity
and low-peaked FSRQs were added, it might appear as more of an ``L''.
With the advent of the {\em Fermi Telescope} era, it
is now possible to also characterize the Compton peak frequency
$\nu_{pk}^C$ and luminosity $L_{pk}^C$.  As we show, the Compton
dominance $A_C\equiv L_{pk}^{C} / L_{pk}^{sy}$ is an important
parameter for charcterizing this sequence.

\section{The 2LAC Blazar Sequence}
\label{2lac_sequence}

The 2LAC \citep{ackermann11_2lac} allows for the characterization of
the high energy component for a greater number of blazars than
previously possible.  Here we look at the blazar sequence among the
2LAC clean sample, which includes 885 total sources, with 395 BL Lacs,
310 FSRQs, and 156 sources of unknown type.

\begin{figure}
%\vspace{2.2mm} 
%\epsscale{1.0} 
%\plotone{fossati}
\includegraphics[width=65mm]{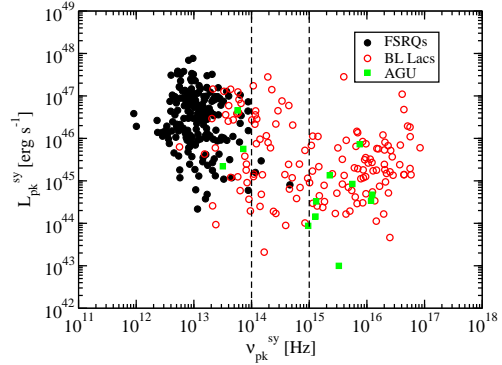}
\caption{Peak synchrotron luminosity versus peak synchrotron frequency
for objects in the 2LAC clean sample.  Filled circles represent FSRQs,
empty circles represent BL Lacs, and filled squares represent objects
which do not have an unambiguous classification.  Dashed lines
indicate the boundary between HSPs and ISPs and between ISPs and LSPs.
}
\label{fossati}
\vspace{2.2mm}
\end{figure}

\cite{abdo10_sed} fit the broadband SEDs of the blazars in the 3-month
LAT bright AGN sample \cite[LBAS;][]{abdo09_lbas} with third degree
polynomials to determine the peak synchrotron frequency,
$\nu_{pk}^{sy}$.  \cite{abdo10_sed} found empirical relations for
finding the peak frequency of the synchrotron component from the slope
between the 5 GHz and 5000~\AA\ flux ($\alpha_{ro}$), and between the
5000~\AA\ and 1 keV flux ($\alpha_{ox}$).  \cite{ackermann11_2lac}
used these empirical relations and their low energy data to determine
$\nu_{pk}^{sy}$ for the 2LAC sample.  This was then used to classify
the SEDs of the blazars as LSP, ISP, or HSP.  \cite{abdo10_sed} also
provided an empirical formula for determining the flux at the
synchrotron peak from the the 5 GHz flux density and $\nu_{pk}^{sy}$.
We used this relation, along with the $\nu_{pk}^{sy}$ values from the
2LAC \citep{ackermann11_2lac} to get the luminosity distance and create
a plot of the peak synchrotron luminosity, $L^{sy}_{pk}$ versus
$\nu_{pk}^{sy}$.  This is plotted in Fig.\ \ref{fossati}.  These are
the sources in the 2LAC clean sample with measured redshifts and
enough SED measurements to determine $\nu_{pk}^{sy}$, which amounts to
352 sources including 145 BL Lacs, 195 FSRQs, and 12 AGN of unknown
optical spectral type (AGUs; i.e., unknown whether they are FSRQs or
BL Lacs).  Note that $\nu_{pk}^{sy}$ is corrected for redshift, and is
in the frame of the source.  We include the photometric redshifts which 
have been determined from \cite{rau12}.

For this diagram, we have computed the Spearman ($\rho$) and Kendall
($\tau$) rank correlation coefficients and the probability of no
correlation (PNC) calculated from each coefficient.  The results can
be found in Table \ref{table_correlate}.  The results from $\rho$ and
$\tau$ are similar in all cases.  The PNC is very small for the BL
Lacs and FSRQs separately, and even lower for all the sources
combined, where the probability is essentially zero that there is not
a correlation.  Note however, that, sources with unknown $z$ are not
included.  This could explain the anti-correlations for the whole
sample and for the BL Lacs in general \cite{giommi12_selection}, since
they could fill in the upper right part of this diagram, as mentioned
in Section \ref{intro}.  However, the objects with unknown $z$ are
almost certainly BL Lacs, so it would not explain any possible
anti-correlation amoung the FSRQs alone, although no significant one
was found.

%\clearpage

%\clearpage
\begin{table*}
\begin{center}
\caption{Statistics of correlations involving $\nu_{pk}^{sy}$.  }
\begin{tabular}{|l|c|c|c|c|}
\hline
Sample & 
$\rho$ & 
PNC($\rho$) & 
$\tau$ & 
PNC($\tau$) \\
\hline
\multicolumn{5}{|c|}{$L_{pk}^{sy}$ versus $\nu_{pk}^{sy}$}  \\
\hline
BL Lacs & -0.19 & 0.019 & -0.12 & 0.035 \\
FSRQs & -0.12 & 0.073 & -0.088 & 0.066 \\
All sources with known $z$ & -0.54 & $4.2\times10^{-28}$ & -0.35 & 0.00 \\
\hline
\multicolumn{5}{|c|}{$L_{5GHz}$ versus $\nu_{pk}^{sy}$}  \\
\hline
BL Lacs & -0.64 & $5.4\times10^{-18}$ & -0.45 & 0.00 \\
FSRQs & -0.36 & $1.82\times10^{-7}$ & -0.25 & $1.79\times10^{-7}$ \\
All sources with known $z$ & -0.79 & 0.00 & -0.58 & 0.00 \\
\hline
\multicolumn{5}{|c|}{$A_C$ versus $\nu_{pk}^{sy}$}  \\
\hline

BL Lacs & -0.30 & $2.7\times10^{-4}$ & -0.21 & $1.3\times10^{-4}$ \\
FSRQs & $8.9\times10^{-3}$ & 0.90 & $6.3\times10^{-3}$ & 0.89 \\
All sources with known $z$ & -0.66 & $9.8\times10^{-45}$ & -0.45 & 0.00 \\
All sources 1 & -0.66 & 0.00 & -0.46 & 0.00 \\
All sources 2 & -0.66 & 0.00 & -0.46 & 0.00 \\
All sources 3 & -0.64 & 0.00 & -0.44 & 0.00 \\
\hline
\end{tabular}
\label{table_correlate}
\end{center}
\end{table*}
%\clearpage

\begin{figure}
%\vspace{2.2mm}
%\epsscale{1.0}
%\plotone{fossati_all_radio}
\includegraphics[width=65mm]{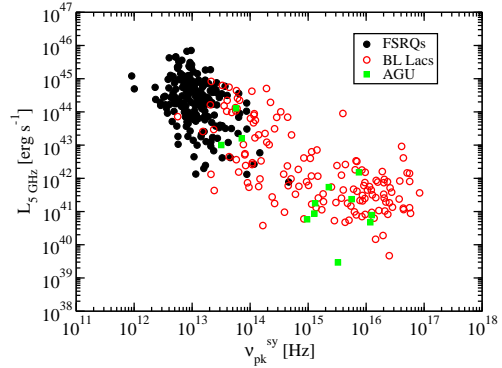}
\caption{Radio luminosity at 5~GHz versus peak synchrotron frequency for 
the 2LAC clean sample.  Symbols are the same as in Fig.\ \ref{fossati}.  
}
\label{fossati_radio}
%\vspace{2.2mm}
\end{figure}

\citep{fossati98} found a more significant anti-correlation between
the 5~GHz luminosity, $L_{5\ GHz}$, and $\nu_{pk}^{sy}$.  We plot this
for our sample in Fig.\ \ref{fossati_radio}.  Like \citep{fossati98},
we find the anti-correlation with $L_{5\ GHz}$ versus $\nu_{pk}^{sy}$
to be much more visually apparent than the one between $L_{pk}^{sy}$
and $\nu_{pk}^{sy}$.  This is also reflected in the decreased PNC for
this diagram (Table \ref{table_correlate}) for both FSRQs and BL Lacs
alone.  If the anti-correlation is explained by the increasing cooling
at higher luminosities \citep{ghisellini98}, the correlation with
$L^{sy}_{pk}$ should be more significant, since the emission at 5 GHz is
thought to be from a different region of the jet than the emission at
the peak.  However, since $\nu_{pk}^{sy}$ was determined in part based
on the 5 GHz flux, this is probably a result of this dependence.  As
pointed our by \citep{lister11}, if $\nu_{pk}^{sy}$ increases, but the
synchrotron bump remains unchanged in other aspects, the radio flux
(or luminosity) will naturally decrease.  This can explain this
anti-correlation.  However, what is unclear is whether a
$\nu_{pk}^{sy}$ derived from the radio flux should be interpreted as a
cooling break \citep{ghisellini98}, since these two should be from
different regions and possibly independent.  Determination of
$\nu_{pk}^{sy}$ independent of low radio frequency emission should be
a good, although technically challenging, way to test this.

%\clearpage

\citep{abdo10_sed} also fit the high energy components of LBAS blazars
with a third degree polynomial to determine the peak of the $\g$-ray
component (presumably from Compton scattering).  They found an
empirical relation between $\nu_{pk}^{C}$ and the LAT $\g$-ray
spectral index, $\G_\g$.  We use this relation to determine the peak
of the Compton component.  Approximately 10\% of the 352 sources are
also found in the 58-month {\em Swift} Burst Alert Telescope (BAT)
catalog.  For these sources, we extrapolated their BAT and LAT
power-laws and found where they intersected.  If they intersected
within the range 195 keV to 100 MeV we used this location as
$\nu_{pk}^C$.  This allows an improved estimation of $\nu_{pk}^C$ over
the empirical relation, particularly for those very soft sources, for
which this empirical relation is untested.  For other sources, the
peak was determined from the LAT spectral index and the empirical
relation from \citep{abdo10_sed}.  Once the location of $\nu_{pk}^C$
is known, either by using the empirical relation or from the BAT-LAT
intersection, the luminosity at the peak, $L_{pk}^{C}$, can be
estimated by extrapolating the LAT spectral index.

\begin{figure}
%\vspace{2.2mm}
%\epsscale{1.0}
%\plotone{CD_bat}
\includegraphics[width=65mm]{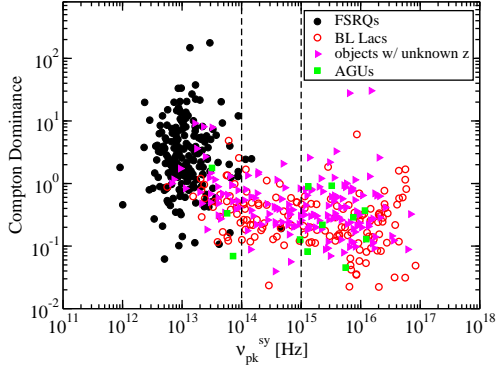}
\caption{Compton dominance (i.e., $L_{pk}^C/L_{pk}^{sy}$)
versus peak synchrotron frequency.  Symbols are the same as in Figures
\ref{fossati} and \ref{fossati_radio}; additionally,
rightward-pointing triangles represent BL Lacs with unknown redshifts,
for which $\nu_{pk}^{sy}$ is a lower limit.  }
\label{CD}
\end{figure}

Combining their results with EGRET data, \cite{fossati98} made a plot
of $\g$-ray dominance versus $\nu_{pk}^{sy}$.  Using LAT data from the
2LAC, as described above, we make a similar plot, although we use
$L_{pk}^C$ instead of simply the $\g$-ray luminosity.  Our results are
in Fig.\ \ref{CD}.  Also note that $A_C$ is independent of redshift.
Thus we can plot in Fig.\ \ref{CD} an additional 174 sources from the
2LAC clean sample that have well-determined synchrotron bumps but do
not have known redshifts.  For these sources, the plotted
$\nu_{pk}^{sy}$ is a lower limit, since the redshifts are not known.
However, it will be larger by only a factor $(1+z)$, i.e., a factor of
a few.

We have also computed the correlation coefficients $\rho$ and $\tau$
for $A_C$ versus $\nu_{pk}^{sy}$, and the results can be found in
Table \ref{table_correlate}.  There is no evidence for a correlation
for the FSRQs alone, although the probability that there is no
correlation for the BL Lacs alone is very small.  For the combined
sample of all sources with known $z$, there is essentially zero chance
that there is no correlation, similar to the $L_{pk}^{sy}$ versus
$\nu_{pk}^{sy}$ correlation.  We also computed the coefficients for
all sources, including the ones with unknown $z$, computing their
$\nu_{pk}^{sy}$ assuming $z=0$ (``all sources 1'' in Table
\ref{table_correlate}).  We find essentially no chance that the
addition of sources with unknown $z$ will explain the correlation when
these sources are included.  The term $\nu_{pk}^{sy}$ will vary by a
factor of a few due to redshift, so we also calculated the
coefficients assuming all these sources with unknown $z$ are at
$z=0.35$, the average of the BL Lacs with known $z$ (``all sources
2''); and assuming these sources are at $z=4$ (``all sources 3''),
which is higher than the maximum redshift of the entire
sample (which is $z=3.1$).  In each case, there is essentially a 100\%
chance that an anti-correlation exists.  Although the objects without
known redshifts could explain the correlation between $L_{pk}^{sy}$ and
$\nu_{pk}^{sy}$, it does not appear they can explain the correlation
between $A_C$ and $\nu_{pk}^{sy}$.  This aspect of the blazar
sequence seems secure.

The two sources with unknown redshifts in the upper right quadrant of
this diagram are 2FGL~J0059.2-0151 (1RXS~005916.3-015030) and
2FGL~J0912.5+2758 (1RXS~J091211.9+27595) with LAT spectral indices of
$\G_\g=1.15\pm0.36$ and $\G_\g=1.20\pm0.37$, respectively.  These are
the two hardest sources in the 2LAC, and the sources with the largest
error bars on their spectral index; only one source in the 2LAC clean
sample is fainter than these sources (2FGL~J1023.6+2959).  They are
clearly outliers.  Propagating the error on their spectral indices,
one finds that they have Compton dominances of $\log_{10}(A_C) =
1.44\pm 1.82$ and $\log_{10}(A_C) = 1.48\pm1.98$, respectively; they
have $A_C$ consistent with unity within their error bars, and so are
consistent with the ``L'' shape seen in Fig.\ \ref{CD}.

% in make_three_pram.pro, these two outliers are:
% object 3:  2FGL~J0059.2-0151  
% sp index = 1.15\pm0.36
% object 92: 2FGL~J0912.5+2758
% sp index = 1.20\pm0.37
%

%\vspace{2.2mm}
%\clearpage

\section{Theoretical Blazar Sequence}
\label{theory}
We describe a simple model for blazar jet emission.  This model is 
similar to the one presented by \cite{boett02_seq}.  We assume the
relativistic jet is dominated by emission from a single zone which is
spherical with radius $R^{\p}_b$ in its comoving frame, and moving
with high relativistic speed $\beta c$ giving it a Lorentz factor
$\G=(1-\beta)^{-1/2}$.  The jet makes an angle to the line of sight
$\theta = \arccos(\mu)$ so that the Doppler factor is
$\dD=[\G(1-\beta\mu)]^{-1}$.  Electrons are injected with a power-law
distribution given by $ Q_e(\g) = Q_0 \g^{-q}\ H(\g;\g_1,\g_2) $.  The
hard X-ray spectra in some blazars may indicate very hard electron
spectra at lower energies \citep{sikora09}.  Although blazars can be
quite variable on timescales as short as hours
\citep[e.g.,][]{abdo09_3c454.3} or even minutes
\citep{aharonian07_2155}, we will assume their average or quiescent
emission can be described by a steady state solution to the electron
continuity equation, where continuous injection is balanced by cooling
and escape.  We assume an energy-independent escape timescale given by
$ t_{esc} = \eta R_b^\prime/c $ where $R_b^\prime$ is the comoving
size of the blob and $\eta$ is a constant $> 1$.  In this case, where
$\g_1 < \g_c$ (the {\em slow-cooling regime}) the electron
distribution can be approximated as
\begin{equation}
N_e(\g) \approx Q_0 t_{esc} \left\{ \begin{array}{ll}
  (\g/\g_c)^{-q}   & \g_1 < \g < \g_c \\
  (\g/\g_c)^{-q-1} & \g_c < \g < \g_2 
      \end{array}
\right. \ . 
\end{equation}
If $\g_c < \g_1$, i.e.,  the {\em fast-cooling regime}, 
\begin{equation}
N_e(\g) \approx Q_0 t_{esc} \left\{ \begin{array}{ll}
  (\g/\g_1)^{-2}   & \g_c < \g < \g_1 \\
  (\g/\g_1)^{-q-1} & \g_1 < \g < \g_2
      \end{array}
\right. \ .
\end{equation}
Here we assume the electrons are cooled by synchrotron emission and
Thomson scattering, so that
\begin{eqnarray}
\label{gc}
\g_c = \frac{ 3 m_e c^2}{ 4c \sT ( \up_B + \up_{sy} + \G^2u_{ext} ) t_{esc} }\ 
\end{eqnarray}
is the cooling electron Lorentz factor, where primes denote quantities
in the comoving frame of the blob.  Here $\up_B = B^2/(8\pi)$ is the
magnetic field energy density, $\up_{sy,tot}$ is the total synchrotron
energy density, and the $u_{ext}$ is the external energy density,
assumed to be isotropic in the proper frame of the AGN.  The exact
nature of the external radiation field is not known, and may not even
be the same for all blazars.  For a given set of parameters, the
nonlinear nature of $\up_{sy}$ means that there is not a simple closed
form solution for $N_e(\g)$, and so we solve for $N_e(\g)$
numerically.

Once $N_e(\g)$ is known, we use the standard formulae in the
$\delta$-approximation
\citep[e.g.,][]{dermer02,finke08_SSC,dermer09_book} to calculate the
synchrotron, SSC, and EC luminosity and energy at their peaks.  This
assumes the Compton-scattering takes place only in the Thomson regime.
We have compared our results with those using the full synchrotron
emissivity and Compton cross section, and found good agreement.  The
Compton dominance ($A_C$) is given by the ratio of the peak
Compton-scattered component to the peak of the synchrotron component,
$ A_C \equiv \max[L^{EC}_{pk},\ L^{SSC}_{pk}]/L^{sy}_{pk} \approx
\max[ \dD^2 u_{ext},\ \up_{sy,pk}]/\up_B\ , $ where we have ignored a
bolometric correction term $\sim 1$.  Note that SSC emission has the
same beaming pattern as synchrotron, and thus
$L^{SSC}_{pk}/L^{sy}_{pk}$ does not depend on the viewing angle;
however, $L^{EC}_{pk}$ does not, and thus $L^{EC}_{pk}/L^{sy}_{pk}$ is
dependent of the viewing angle through $\dD$
\citep{dermer95,georgan01}.

We found we could adequately reproduce the sources in Figures
\ref{fossati} and \ref{CD} without changing the power injected in
electrons.  Instead, we only vary $B$ and $u_{ext}$, assuming
$B\propto u_{ext}^{1/4}$, and the angle to the line of sight,
$\theta$.  Further details can be found in \cite{finke12}.

\begin{figure}
%\epsscale{1.0}
%\plotone{fossati_06}
\includegraphics[width=65mm]{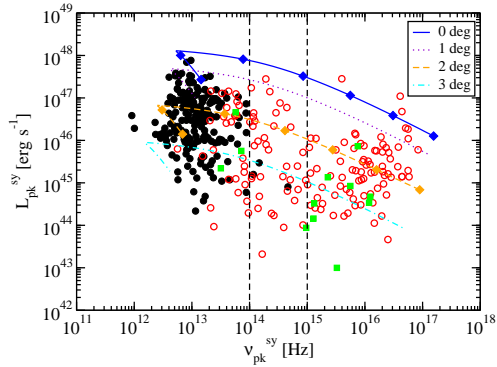}
\caption{
Same as Figure \ref{fossati}, with curves showing our
model plotted at various angles, shown in the legend.  Along the
curves only $B$ and $u_{ext}$ are varyed.  
}
\label{fossati_theory}
\end{figure}

Using the model described above, we reproduce the $L_{pk}^{sy}$ versus
$\nu_{pk}^{sy}$ relation as seen in the curves in Fig.\
\ref{fossati_theory}.  At $\lesssim 10^{13}$\ Hz (the exact value depends on
$\theta$), $\g_c$ will become less than $\g_1$, leading to a sharp
inflection.  By changing the viewing angle, the synchrotron luminosity
decreases dramatically, so that almost all of the BL Lacs can be
reproduced by this model.  Note that this model predicts that there
will be sources found with $L_{pk}^{sy}\gtrsim10^{47}\ \erg\ \s^{-1}$ and
$\nu_{pk}^{sy}\gtrsim10^{15}\ \Hz$, which is relatively unpopulated.  This
region could be filled in by bright BL Lacs at high $z$, which for
which we cannot determine a redshift due to the nonthermal emission
washing out their weak emission lines \cite{giommi12_selection}.
Indeed, this region is beginning to be filled in by constraining the
redshifts of several BL Lacs \citep{padovani12}.

%\clearpage

\begin{figure}
\vspace{3.6mm} 
%\epsscale{1.0}
%\plotone{CD_08}
\includegraphics[width=65mm]{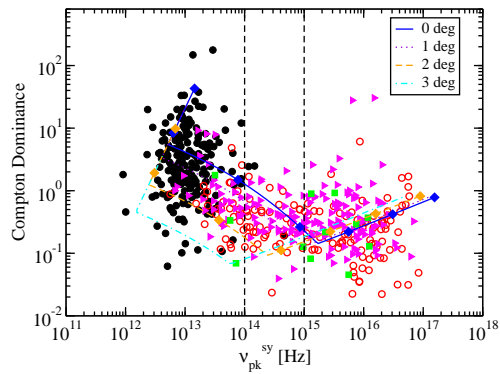}
\caption{ Same as Figure \ref{CD} with curves showing our model
plotted at various angles, shown in the legend.  Along the curves only
$B$ and $u_{ext}$ are varyed.  }

\label{CD_theory}
\end{figure}

The model curves for $A_C$ versus $\nu_{pk}^{sy}$ are plotted in Fig.\
\ref{CD_theory}.  Again, the model seems to reproduce the 2LAC data
well.  There is a distinct difference above and below the transition
between the Compton component being dominated by SSC and being
dominated by EC, at around $\nu_{pk}^{sy}\approx2\times10^{15}\ \Hz$
for the $\theta=0$ curve.  Note that $A_C$ is not dependent on
$\theta$ for SSC, although the curve here does shift to lower
$\nu_{pk}^{sy}$ due to its dependence on $\theta$. The EC part of the
curves, however, are strongly dependent on $\theta$ through $\dD$.  As
with Fig.\ \ref{fossati_theory}, a sharp inflection is seen at lower
frequencies ($\nu_{pk}^{sy}\approx 10^{13}\ \Hz$ for $\theta=0$) due
to the transition of the peak from being associated with $\g_c$ to
$\g_1$.  Also as with the previous figure, the BL Lacs are
well-reproduced, while the FSRQs are not, particularly those with high
$A_C$.  A diversity of $\g_1$ would clearly allow a widening of this
branch, and allow the model to reproduce all of the FSRQs.
Alternatively, if the jets of blazars are made up of multiple emitting
components with different Lorentz factors, so that at different angles
a different jet component would dominate the SED, this could also
broaden the sequence and reproduce more objects \citep{meyer11}.

%\clearpage

\section{Discussion}

For blazars, it has been suggested that the peak of the electron
distribution ($\gamma_{pk}$) is independent of broad emission line
luminosities, $B$, and other properties \citep{giommi12_selection}.
If this were the case, and if the broad lines are the source for
external scattering, sources would be found with high $A_C$
($A_C\gtrsim$ a few) and and high $\nu_{pk}^{sy}$
($\nu_{pk}^{sy}\gtrsim 10^{15}\ \Hz$).  The redshift is not relevant,
since both of these quantities are essentially redshift independent.
However, no such sources are found in our sample, including BL Lacs
without known redshift (Fig. \ref{CD_theory}).  If $\gamma$ rays are
produced by external Compton processes, as seems likely
\citep{meyer12}, there does indeed seem to be a relation between
$u_{ext}$ and $\gamma_{pk}$.  We have found a simple model that
explains this, where $\gamma_{pk}=\max[\gamma_c, \gamma_1]$, where
$\gamma_c$ is dependent on both $u_{ext}$ and $B$ (Equation
[\ref{gc}]).  This is quite similar to previous work
\citep[e.g.,][]{ghisellini98,boett02_seq}, the main different being
the dependence of $B$ and $u_{ext}$.  This simple model predicts
sources will be found with $L_{pk}^{sy}\gtrsim 10^{47}\ \erg\ \s^{-1}$
and $\nu_{pk}^{sy}\gtrsim 10^{15}\ \Hz$ (Fig. \ref{fossati_theory}),
when the redshifts of more BL Lacs are determined
\citep[unlike][]{ghisellini98,boett02_seq}.  These sources should be
brighter because they are more highly aligned with our line of sight,
rather than intrinsic brightness.  \cite{ghisellini12_seq} also
predict that sources will be found in this region, although in their
case these sources will be ``blue FSRQs'', with the primary emitting 
region ourside the BLR, assuming in other FSRQs, broad line photons 
are the external radiation source for Compton scattering.

%$\star -$

%$\smile$

%$\bowtie$

\begin{acknowledgments}

We are grateful to M.\ Lister, M.\ Georganopoulos, and K.\ Wood for
useful discussions on the blazar sequence.  We are also grateful to
R.\ Ojha and all of the organizers for a very interesting and
well-organized conference.  This work was partially supported by {\em
Fermi} GI Grant NNH09ZDA001N.

\end{acknowledgments}

\bigskip % extra skip inserted
\bibliography{3c454.3_ref,EBL_ref,references,mypapers_ref,sequence_ref,blazar_ref,SSC_ref}

\end{document}